\documentstyle[prl,aps,psfig]{revtex}
\begin{document}

\title{ Fluctuation Conductivity in Mesoscopic Superconductor -
 Normal Metal Contacts }

\author{ A.\ F.\ Volkov$^{* +}$ , K.\ E.\ Nagaev$^*$  and  R.\ Seviour$^+$}

\address{* Institute of Radioengineering and Electronics, Russian Academy of
Sciences, Mokhovaya ul. 11, Moscow 103907, Russia. \\ + School of Physics and
Chemistry, Lancaster University, Lancaster LA1 4YB, U.K.}

\maketitle

\begin{abstract}
The fluctuation Conduction of NSN and SNS contacts above $T_c$
are analyzed. For NSN contacts, both Aslamazov - Larkin and Maki
- Thompson corrections to the conduction are found to be of the same order
and diverge for $T_c^* < T_c$ according to the law $(T -
T_c^*)^{-1}$. For SNS contacts, the Aslamazov - Larkin correction
vanishes, while the Maki - Thompson correction is essential for
contacts shorter than the phase-breaking length.
\end{abstract}

\bigskip
Pacs numbers: 72.10Bg, 73.40Gk, 74.50.+r

\section{INTRODUCTION}

Recently, mesoscopic superconducting - normal metal(S/N) systems
have attracted a great deal of attention \cite{Petrashov} - \cite{Nazarov}.
 In particular, it was shown that their conductance exhibits 
oscillatory behavior in magnetic field \cite{Petrashov} - 
\cite{Courtois} and nonmonotonic temperature and voltage 
dependences \cite{Courtois}. The reason for this behaviour is the 
effect superconducting correlations have on the electrons in the
normal metal. The physics of these effects is simillar to
 the physics of corrections to the conductivity resulting 
from superconducting fluctuations above $T_c$ \cite{Aslamazov},
\cite{Maki}, \cite{Thompson}. In particular, the authors of ref
\cite{Moriond} have shown that the nonmonotonic temperature dependence 
of conductance in S/N systems is the result of competition 
between the contribution from the modified density of states 
and a contribution which is similar to the Maki - Thompson (MT)
fluctuation conductivity above $T_c$ \cite{Maki}, \cite{Thompson}. 
Therefore, it is of interest to calculate the conductance
of different S/N systems above $T_c$ taking into account superconducting 
fluctuations.

Despite the large number of papers concerned with superconducting
fluctuations in macroscopic samples, very few authors have considered
superconducting fluctuations in contacts. In particular, Kulik
\cite{Kulik} considered the effect of superconducting
fluctuations on the density of states and on the current in a
tunnel junction. Zaitsev \cite{Zaitsev-84} considered the
fluctuation corrections to the conductance of very short
superconducting microbridges. However, these studies have revealed
only two types of fluctuation corrections in uniform
systems; the correction due to the modified density of states and the MT
correction, which represents the effect of fluctuational Cooper
pairs on the conduction of normal electrons. They did not reveal
the Aslamazov - Larkin (AL) correction \cite{Aslamazov} which
represents the direct contribution of fluctuational Cooper pairs
to the current.

In this paper, we consider the effects of superconducting
fluctuations on the conductance of mesoscopic NSN and SNS
contacts of various lengths. In the case of NSN
contacts, we find that the AL and MT corrections are of the same
order of magnitude. In the case of SNS contacts, the conductance
is determined by the MT correction, which penetrates into the
contact, from the electrodes, a distance upto the
phase-breaking length $L_{\varphi}$.

\section{ BASIC EQUATIONS }

The expressions determining the superconducting corrections to
the conductivity are easily obtained by a trivial extension of
the Aslamazov - Larkin and Maki - Thompson equations to
inhomogeneous systems. However, as many people are not familiar
with the diagrammatic technique used by these authors, we
present here a different derivation based on quasiclassical
Green's functions of the superconductor and the self-consistency
equation with a Langevin source \cite{Larkin-73},\cite{Zaitsev-84},
\cite{Volkov-93}. One can show that the results obtained with the 
aid of the diagrammatic technique and the results presented here
are identical.

The current density in a dirty superconductor is expressed by
the formula
 \begin{equation}
 {\bf j} 
 =
 \frac{\pi}{2} e N_F D
 \int\frac{d\epsilon}{2\pi}
 \int\frac{d\epsilon '}{2\pi}\,
 {\rm Sp}
 \left\{
   \hat{\tau}_z
   \left[
      \hat{g}^R(\epsilon, \epsilon ')
      \frac{\partial\hat{g}^F(\epsilon, \epsilon ')}{\partial\bf r} 
      -
      \hat{g}^F(\epsilon, \epsilon ')
      \frac{\partial\hat{g}^A(\epsilon, \epsilon ')}{\partial\bf r} 
   \right]
 \right\},
 \label{j-basic}
 \end{equation}
where $\hat{g}^R$, $\hat{g}^A$, and $\hat{g}^F$ are
quasiclassical matrix Green's functions of the superconductor
\cite{Larkin-86}, $N_F = mp_0/2\pi^2$ is the density of states
at the Fermi level, and $D = lv_0/3$ is the diffusion
coefficient of electrons. The retarded and advanced Green's
functions $\hat{g}^{R(A)}$ describe the energy spectrum of the
superconductor, while $\hat{g}^F$ also contains information
about the electron distribution. In the case of a time-independent
electrical potential, the functions $\hat{g}^{R(A)}(\epsilon,
\epsilon ', {\bf r})$ obey the equation
 \begin{equation}
 -D\frac{\partial}{\partial\bf r}
 \left(
   \hat{g}^{R(A)}\frac{\partial\hat{g}^{R(A)}}{\partial\bf r}
 \right)
 +
 \left(
  -i\epsilon\hat{\tau}_z - i\hat{\Delta} + \hat{\Sigma}_d^{R(A)} 
 \right)
 \hat{g}^{R(A)}
 -
 \hat{g}^{R(A)}
 \left(
  -i\epsilon'\hat{\tau}_z - i\hat{\Delta} + \hat{\Sigma}_d^{R(A)} 
 \right)
 =
 0
 \label{gR-gA}
 \end{equation}
where
 $$
 \hat{\Delta}
 =
 \left(
 \begin{array}{ccc}   
   0            & &     \Delta  \\
   -\Delta^*    & &     0       \\
 \end{array}
 \right),
 $$
and the matrix $\hat{\Sigma}_d^{R(A)} = \pm\Gamma\hat{\tau}_z$
describes the depairing. The products of the matrix quantities also
imply convolutions over the inner frequencies.

The order parameter $\Delta(\omega, {\bf r})$ satisfies the
self-consistency equation containing the source of condensate fluctuations
\cite{Larkin-73}:
 \begin{equation}
 \Delta(\omega, {\bf r})
 =
 \frac{\lambda}{8}
 \int
 \frac{d\epsilon}{2\pi}
 {\rm Sp}
 \left[
    (\hat{\tau}_x - \hat{\tau}_y)
    \hat{g}^F(\epsilon + \omega/2, \epsilon - \omega/2, {\bf r})
 \right]
 -
 \lambda\eta(\omega, {\bf r}),
 \label{self-consistency}
 \end{equation}
and the correlation function of the sources of fluctuations $\eta$ is
given by
 \begin{eqnarray}
 &
 <\eta(\omega, {\bf r})\eta^*(\omega ', {\bf r'})>
 =
 (16\pi N_F)^{-1}
 \delta(\omega + \omega ')
 \delta({\bf r - r'})
 &
             \nonumber\\
 &
 <\eta(\omega, {\bf r})\eta(\omega ', {\bf r'})>
 =
 0.
 & 
 \label{correlator}
 \end{eqnarray}

This expression implies that the sources of condenstate
fluctuations are $\delta-$ correlated in time. The latter 
equality is the result of
randomness in the phase of superconducting fluctuations above
$T_c$. 

Since the fluctuations of the order parameter are small, the
retarded and advanced Green's functions of the superconductor
may be represented as the sum of corresponding normal-metal Green's
functions and a small additive proportional to $\Delta$:
 \begin{equation}
 \hat{g}^{R(A)}(\epsilon, \epsilon ')
 =
 \pm 2\pi\hat{\tau}_z \delta(\epsilon - \epsilon ')
 +
 \hat{f}^{R(A)}(\epsilon, \epsilon ').
 \label{substitute}
 \end{equation}
Substituting Eqn. (\ref{substitute}) into Eqn. (\ref{gR-gA}) and
making use of the orthogonality condition $[\hat{g}^{R(A)}]^2 =
\hat{1}$ \cite{Larkin-86}, one obtains that
 \begin{equation}
 \hat{f}^{R(A)}(\epsilon, \epsilon ', {\bf r})
 =
 \pm\int d^3{\bf r'}
 \,
 P^{R(A)}(\epsilon + \epsilon ', {\bf r}, {\bf r'})
 \,
 \hat{\Delta}(\epsilon - \epsilon ', {\bf r'}),
 \label{fR-explicit}
 \end{equation}
where the kernels $P^{R(A)}$ are determined by the equations 
 \begin{eqnarray}
 &
 \left(
   D\partial^2/\partial {\bf r}^2
   +
   i\epsilon - \Gamma
 \right)
 P^R(\epsilon, {\bf r}, {\bf r'})
 =
 2i\delta({\bf r - r'})
 &
 \nonumber\\
 &
 P^A(\epsilon, {\bf r}, {\bf r'})
 =
 -P^R(-\epsilon, {\bf r}, {\bf r'}).
 &
 \label{PR-definition}
 \end{eqnarray}

Note that the $\hat{f}^{R(A)}$ matrices contain no diagonal
components. The correction to the diagonal components, which
determines the density of states, is proportional to the order
parameter squared. However, this correction is small for the
case under consideration and will be neglected by us.

As the local electron distribution is assumed to be 
equilibrium, the function $\hat{g}^F$ may be expressed in terms of 
$\hat{g}^R$ and $\hat{g}^A$ via the relationship
 \begin{equation}
 \hat{g}^F(\epsilon, \epsilon ')
 =
 \hat{g}^R(\epsilon, \epsilon ')\,
 \hat{n}(\epsilon ')
 -
 \hat{n}(\epsilon)
 \hat{g}^A(\epsilon, \epsilon '),
 \label{gF}
 \end{equation}
where
 $$
 \hat{n}(\epsilon, {\bf r})
 = 
 \left(
   \begin{array}{cc}
   n(\epsilon - e\phi({\bf r}))  &   0                          \\
   0                            &  n(\epsilon + e\phi({\bf r})) \\
   \end{array}
 \right),
 $$
 \begin{equation}
 n(\epsilon) = \tanh(\epsilon/2T).
 \label{n-definition}
 \end{equation}
and $\phi({\bf r})$ is the electric potential. Substituting Eqns. (\ref{gF}), (\ref{fR-explicit}), and
(\ref{substitute}) into the self-consistency equation
(\ref{self-consistency}) gives
 \begin{eqnarray}
 \Delta(\omega, {\bf r})
 =
 \frac{\lambda}{4}
 \int d^3 {\bf r'}
 \int \frac{d\epsilon}{2\pi}
 \Bigl[
    P^R(2\epsilon, {\bf r}, {\bf r'})
    & &
    n(\epsilon - \omega/2 + e\phi({\bf r}))
 \nonumber\\
    +
    & &
    P^A(2\epsilon, {\bf r}, {\bf r'})
    n(\epsilon + \omega/2 - e\phi({\bf r}))
 \Bigr]
 \Delta(\omega, {\bf r'})
 -
 \lambda\eta(\omega, {\bf r}).
 \label{self-consistency-2}
 \end{eqnarray}

Consider the case where all the relevant length scales are much
larger than the characteristic length $\xi_0 \sim (D/T_c)^{1/2}$
and $\omega \ll T_c$. Then $\Delta({\bf r'})$ may be expanded in
powers of ${\bf r - r'}$ to quadratic terms. Substituting the
expression for $P^{R(A)}$ in infinite space,
 \begin{equation}
 P_0^{R(A)}(2\epsilon, {\bf r - r'})
 =
 \int\frac{d^3 q}{(2\pi)^3}
 \,
 \frac
 {
  \exp[i{\bf q}({\bf r - r'})]
 }
 {
  \epsilon \pm i(Dq^2 + \Gamma)/2
 }, 
 \label{PR0}
 \end{equation}
into the self-consistency equation (\ref{self-consistency-2}),
one obtains the nonstationary Ginzburg - Landau - Langevin equation in the
form 
 \begin{equation}
 \left(
    D\frac{\partial^2}{\partial {\bf r}^2}
    +
    i\omega
    -
    2ie\phi({\bf r})
    -
    \tau T_c
    -
    \Gamma
 \right)
 \Delta(\omega, {\bf r})
 =
 16T\eta(\omega, {\bf r}),
 \label{Ginzburg}
 \end{equation}
where $T_c$ is the BCS transition temperature and $\tau =
(8/\pi)(T - T_c)/T_c$. This equation is well known in the theory
of superconducting fluctuations.

Consider now the expression for the current (\ref{j-basic}).
Substituting Eqns. (\ref{substitute}) and (\ref{gF}),
one obtains:
 \begin{equation}
 {\bf j} = {\bf j}_n + {\bf j}_{AL} + {\bf j}_{MT},
 \label{j-sum}
 \end{equation}
where
 \begin{equation}
 {\bf j}_n
 =
 \frac{1}{2}N_F De\int d\epsilon
 \,
 {\rm Sp}
 \left(
    \hat{\tau}_z
    \frac{\partial\hat{n}}{\partial {\bf r}}
 \right)
 =
 \sigma_n {\bf  E};
 \label{jn}
 \end{equation}
$\sigma_n$ represents the normal-state conduction. The second term represents
the regular Aslamazov - Larkin (AL) correction
 \begin{equation}
 {\bf j}_{AL}
 =
 \frac{\pi}{2} e N_F D
 \int\frac{d\epsilon}{2\pi}
 \int\frac{d\epsilon'}{2\pi}
 \,
 {\rm Sp}
 \biggl\{
     \hat{\tau}_z
     \biggl[
          \hat{f}^R(\epsilon, \epsilon')
          \frac{\partial\hat{f}^R(\epsilon', \epsilon)}
               {\partial {\bf r}}
          \hat{n}(\epsilon)
          -
          \hat{n}(\epsilon)
          \hat{f}^A(\epsilon, \epsilon')
          \frac{\partial\hat{f}^A(\epsilon', \epsilon)}
               {\partial {\bf r}}
     \biggr]
 \biggr\}
 \label{AL1}
 \end{equation}
and the third term represents the anomalous Maki - Thompson (MT)
correction 
 \begin{equation}
  {\bf j}_{MT}
 =
 -\frac{\pi}{2} e N_F D
 \int\frac{d\epsilon}{2\pi}
 \int\frac{d\epsilon'}{2\pi}
 \,
 {\rm Sp}
 \left[
     \hat{\tau}_z
     \hat{f}^R(\epsilon, \epsilon')
     \frac{\partial\hat{n}(\epsilon')}{\partial {\bf r}}
     \hat{f}^A(\epsilon', \epsilon)
 \right].
 \label{MT1}
 \end{equation}

First consider the AL correction. Substituting Eqn.
(\ref{fR-explicit}) into Eqn. (\ref{AL1}) and performing the
averaging over the fluctuations of the order parameter, one
obtains 
 $$
 {\bf j}_{AL}
 =
 \frac{\pi}{2}eN_FD
 \int\frac{d\epsilon}{2\pi}
 \int\frac{d\omega}{2\pi}
 \int d^3r_1
 \int d^3r_2
 \biggl[
    P^R(2\epsilon - \omega, {\bf r}, {\bf r}_1)
    \frac{\partial P^R(2\epsilon - \omega, {\bf r}, {\bf r}_2)}
         {\partial\bf r}
 $$
 $$
    -
    P^A(2\epsilon - \omega, {\bf r}, {\bf r}_1)
    \frac{\partial P^A(2\epsilon - \omega, {\bf r}, {\bf r}_2)}
         {\partial\bf r}
 \biggr]
 \bigl[
    <
      \Delta^*(\omega, {\bf r}_1)
      \Delta  (-\omega, {\bf r}_2)
    >
    n(\epsilon + e\phi)
 $$
 \begin{equation}
  -
    <
      \Delta(\omega, {\bf r}_1)
      \Delta^*(-\omega, {\bf r}_2)
    >  
     n(\epsilon - e\phi)
 \bigr].
 \label{AL2}
 \end{equation}

Equation (\ref{AL2}) may be simplified, when the
characteristic length scales are much larger than $\xi_0$,
 by setting ${\bf r}_1 = {\bf r}$ in the correlators in second
factor of the integrand of (\ref{AL2}) and expanding in powers
of ${\bf r}_2 - {\bf r}$ to linear terms. Making use of Eqn.
(\ref{PR0}) for $P^{R(A)}$ in the infinite space, Eqn.
(\ref{AL2}) is easily shown to be of the form
 \begin{equation}
 {\bf j}_{AL}({\bf r})
 =
 -\frac{i}{4}eN_FD\int d\omega
 \biggr[
     \left<
        \frac{\partial\Delta(\omega, {\bf r})}{\partial\bf r}
        \Delta^*(-\omega, {\bf r})
     \right>
     \left.
         \frac{\partial n}{\partial\epsilon}
     \right|_{-\omega/2 + e\phi({\bf r})}
     -
     \left<
         \Delta(\omega, {\bf r})
         \frac{\partial\Delta^*(-\omega, {\bf r})}{\partial\bf r} 
     \right>
     \left.
         \frac{\partial n}{\partial\epsilon}
     \right|_{\omega/2 - e\phi({\bf r})}
 \biggl].
 \label{AL3}
 \end{equation}

This is just the standard Ginzburg - Landau expression for the
current. Combined with Eqn. (\ref{Ginzburg}), it gives the
correction to the current due to fluctuations, which include 
the effects of a nonlinear electric field.

We now calculate the AL correction to linear terms in the
electric field. Using the functions $K^R(\omega,
{\bf\rho}, {\bf\rho'})$, the Green's function of Eqn.
(\ref{Ginzburg}) 
with zero potential, and $K^A(\omega, {\bf\rho}, {\bf\rho'}) =
K^R(-\omega, {\bf\rho}, {\bf\rho'})$, then retaining only linear 
terms in the electric field, the fluctuation of the order parameter
 may be written in the form
 $$
 \Delta(\omega, {\bf r}) 
 =
 \int d^3 {\bf r'}
 \,
 K^R(\omega, {\bf r}, {\bf r'})\eta(\omega, {\bf r'})
 +
 2ie
 \int d^3 {\bf r'}
 \int d^3 {\bf r}_1
 \,
 K^R(\omega, {\bf r}, {\bf r}_1)
 [
  \phi({\bf r}_1) 
  -
  \phi({\bf r})
 ]
 K^R(\omega, {\bf r}_1, {\bf r'})\eta(\omega, {\bf r'}),
 $$
 \begin{equation}
 \Delta^*(-\omega, {\bf r}) 
 =
 \int d^3 {\bf r'}
 \,
 K^A(\omega, {\bf r}, {\bf r'})\eta((\omega, {\bf r'})
 -
 2ie
 \int d^3 {\bf r'}
 \int d^3 {\bf r}_1
 \,
 K^A(\omega, {\bf r}, {\bf r}_1)
 [
  \phi({\bf r}_1) 
  -
  \phi({\bf r})
 ]
 K^A(\omega, {\bf r}_1, {\bf r'})\eta^*(-\omega, {\bf r'}).
 \label{Delta}
 \end{equation}

Note that in comparison with $K^R(\omega,{\bf\rho},{\bf\rho'})$ defined
by Aslamasov and Larkin, this quantity contains an additional factor $m p_0 / 16 \pi T_c$.
Substituting Eqn. (\ref{Delta}) and the correlator of the
 sources of condensate fluctuations  Eqn. (\ref{correlator}) into Eqn. (\ref{AL3}),
one obtains the linear AL correction in the form
 $$
 {\bf j}_{AL}({\bf r})
 =
 \frac{8}{\pi}e^2DT
 \int d\omega
 \int d^3 {\bf r}_1
 \int d^3 {\bf r}_2
 \frac{\partial K^R({\bf r}, {\bf r}_1)}{\partial {\bf r}}
 \bigl\{
    [
     \phi({\bf r}_1) 
     -
     \phi({\bf r})
    ]
    K^R({\bf r}_1, {\bf r}_2)
 $$
 \begin{equation}
    -
    [
     \phi({\bf r}_2) 
     -
     \phi({\bf r})
    ]
    K^A({\bf r}_2, {\bf r}_1)
 \bigr\}
 K^A({\bf r}, {\bf r}_2).
 \label{AL-basic}
 \end{equation}

Note that for the AL correction, the current - field
relationship is substantially nonlocal, i.e., the current
density at a given point is determined by the electric field in
its vicinity, of radius $\xi(T) \sim \sqrt{D/(T - T_c)}$. This is
a consequence of the large size of fluctuational Cooper pairs near
$T_c$.

Now we proceed to the MT correction. Substituting Eqn.
(\ref{fR-explicit}) for $\hat{f}_{R(A)}$ into Eqn. (\ref{MT1}),
one obtains
 $$
 {\bf j}_{MT}({\bf r})
 =
 -\pi e^2 N_F D\frac{\partial\phi}{\partial\bf r}
 \int d^3 r_1
 \int d^3 r_2
 \int\frac{d\omega}{2\pi}
 \left<
    \Delta(\omega, {\bf r}_1) 
    \Delta^*(-\omega, {\bf r}_2) 
 \right>
 $$
 \begin{equation}
 \times
 \int\frac{d\epsilon}{2\pi}
 P^R(2\epsilon, {\bf r}, {\bf r}_1)
 P^A(2\epsilon, {\bf r}, {\bf r}_2)
 \frac{\partial}{\partial\epsilon}
 n(\epsilon - \omega/2 + e\phi({\bf r})).
 \label{MT2}
 \end{equation}
Retaining only the terms linear in the electric field the MT correction 
is given by the expression
 \begin{equation}
 \sigma_{MT}({\bf r})
 =
 8e^2DT
 \int d^3 r_1
 \int d^3 r_2
 \int d^3 r'
 \int\frac{d\omega}{2\pi}
 K^R(\omega, {\bf r}_1, {\bf r'})
 K^A(\omega, {\bf r}_2, {\bf r'})
 \int\frac{d\epsilon}{2\pi}
 P^R(2\epsilon, {\bf r}, {\bf r}_1)
 P^A(2\epsilon, {\bf r}, {\bf r}_2)
 \label{MT-basic}
 \end{equation}

Unlike the AL correction, the current-field relationship of the
MT correction is purely local, as the electric field
directly affects normal electrons rather than fluctuational
Cooper pairs.

\section{ NSN CONTACT }

Consider the NSN contact in the shape of a narrow channel of
length $L \gg \xi_0$ and cross-sectional area $S$ connecting two
massive electrodes. The transverse dimensions of the channel are
assumed to be much smaller than $\xi(T)$. Let the $x$ axis be
directed along the channel. The normal-state electric potential
in the contact is unperturbed by superconducting fluctuations and has 
the form $\phi = -Vx/L$, where $V$ is the voltage drop across
the contact. As the superconducting corrections to the current
essentially depend on the distance from the electrodes, the AL
and MT corrections (Eqns. (\ref{AL-basic}) , (\ref{MT-basic})) should
be calculated with the unperturbed potential and then averaged
over the length of the contact to ensure current conservation.

First consider the AL correction. As the transverse dimensions
of the channel are small, all the relevant quantities may be
considered to be dependent only on the longitudinal coordinate
$x$. Introduce a system of the eigenfuctions of the Laplace equation
 \begin{equation}
 \varphi_n(x)
 =
 \sqrt{\frac{2}{L}}
 \sin
 \left(
    \frac{\pi nx}{L}
 \right).
 \label{phi-n}
 \end{equation}
Then the function $K^R$ entering into Eqn. (\ref{AL-basic}), the
Green's function of Eqn. (\ref{Ginzburg}) with zero boundary
conditions at the ends of the contact, may be represented in the form
 \begin{equation}
 K^R(x, x')
 =
 -\frac{i}{S}
 \sum_n
 \frac{\varphi_n(x)\varphi_n(x')}
      {\omega + i\epsilon_L(\pi^2n^2 + \tau_L + \gamma)},
 \label{KR-series}
 \end{equation}
where $\epsilon_L = D/L^2$ is the Thouless energy, $\tau_L =
\tau T_c/\epsilon_L$, and $\gamma = \Gamma/\epsilon_L$.
Performing the integration in Eqn. (\ref{AL-basic}) over the 
coordinates and frequency and making use of the relationships
 \begin{equation}
 \int\limits_0^L dx
 \,
 \varphi_m(x) x \varphi_n(x)
 \,
 =
 \,
 \left\{
    \begin{array}{cc}
       -\frac{\displaystyle 4}{\displaystyle\pi^2} L
       \frac{\displaystyle mn}{\displaystyle (m^2 - n^2)^2}
       \bigr[
          1 - (-1)^{m + n}
       \bigl],
       \quad
       &
       m \ne n
       \\ & \\
       L/2,
       &
       m = n,
    \end{array}
 \right.
 \label{x-mn}
 \end{equation}
 \begin{equation}
 \int\limits_0^L dx
 \,
 \frac{\partial\varphi_m}{\partial x} \varphi_n(x)
 =
 \left\{
    \begin{array}{cc}
       -\frac{\displaystyle 2}{\displaystyle L} 
       \frac{\displaystyle mn}{\displaystyle n^2 - m^2}
       \bigr[
          1 - (-1)^{m + n}
       \bigl],
       \quad
       &
       m \ne n
       \\ & \\
       0,
       &
       m = n,
    \end{array}
 \right.
 \label{dx-mn}
 \end{equation}
one obtains the AL correction in the form
 \begin{equation}
 \delta I_{AL}
 =
 64 \frac{e^2 TV}{\epsilon_L}
 \sum_m
 \sum_{n \ne m}
 \left(
    \frac{mn}{m^2 - n^2}
 \right)^2
 \frac
 {
  \bigl[ 
        1 - (-1)^{m + n}
  \bigr]^2
 }
 {
   \theta_m\theta_n(\theta_m + \theta_n)
 },
 \label{AL-series}
 \end{equation}
where $\theta_m = \pi^2m^2 + \tau_L + \gamma$. In the limit
$\tau_L \gg 1 \gg \gamma$, Eqn. (\ref{AL-series}) gives the
standard AL correction for the one-dimensional wire
 \begin{equation}
 \delta I_{AL}
 =
 \frac{\pi^{3/2}}{2^{11/2}}
 \frac{e^2 T\epsilon_L^{1/2}}{(T - T_c)^{3/2}}
 V.
 \label{AL-1D}
 \end{equation}
The AL correction (\ref{AL-series}) remains finite at $T = T_c$
owing to the finite length of the contact, the
transition temperature decreases from the bulk value to the
value 
 \begin{equation}
 T_c^* = T_c - \frac{\pi}{8}\Gamma - \frac{\pi^3}{8}\epsilon_L.
 \label{Tc*}
 \end{equation}
Near $T_c^*$ the temperature dependence of AL correction is of
the form
 $$
 \delta I_{AL}
 =
 a\frac{e^2TV}{T - T_c^*},
 $$
 \begin{equation}
 a 
 =
 \frac{1}{\pi^3}
 \sum_{k=1}^{\infty}
 \frac{k^2}{(k^2 - 1)^4}
 \approx
 0.1026.
 \label{AL-Tc}
 \end{equation}

Now we proceed to the MT correction. As the quantities
$P^{R(A)}$ appearing in Eqn. (\ref{MT-basic}) also satisfy the zero boundary
conditions at the ends of the contacts, they have the form
 \begin{equation}
 P^R (\epsilon, x, x')
 =
 \frac{2}{S}
 \sum_n
 \frac
 {
  \varphi_n(x) \varphi_n(x')
 }
 {
  \epsilon
  +
  i\epsilon_L (\pi^2 n^2 + \gamma)
 }.
 \label{PR-series}
 \end{equation}
Substituting Eqns. (\ref{KR-series}) and (\ref{PR-series})
into Eqn. (\ref{MT-basic}), averaging over the contact
length and summing the resulting series one obtains
 \begin{equation}
 \delta I_{MT}
 =
 2e^2 V\frac{1}{\tau}
 \left\{
    \frac{1}{\gamma^{1/2}}
    \left(
       \coth \gamma^{1/2}
       -
       \frac{1}{\gamma^{1/2}}
    \right)
    -
    \frac{1}{(\tau_L + \gamma)^{1/2}}
    \left[
      \coth (\tau_L + \gamma)^{1/2}
      -
      \frac{1}{(\tau_L + \gamma)^{1/2}}
    \right]
 \right\}
 \label{T>Tc}
 \end{equation}
for $\tau_L + \gamma > 0$ and
 \begin{equation}
 \delta I_{MT}
 =
 2e^2 V\frac{1}{|\tau|}
 \left\{
    \frac{1}{|\tau_L + \gamma|^{1/2}}
    \left[
      \frac{1}{|\tau_L + \gamma|^{1/2}}
      -
      \cot |\tau_L + \gamma|^{1/2}
    \right]
    -
    \frac{1}{\gamma^{1/2}}
    \left(
     \coth \gamma^{1/2}
     -
     \frac{1}{\gamma^{1/2}}
    \right)
 \right\}
 \label{T<Tc}
 \end{equation}
for $\tau_L + \gamma < 0$. Alternatively, Eqns.(\ref{T>Tc}) - (\ref{T<Tc})
 can be written in the form
 \begin{equation}
 \delta I_{MT}
 =
 4 \frac{e^2 TV}{\epsilon_L}
 \sum_m 
 \frac
 {
        1 
 }
 {
   \theta_m(\theta_m - \tau_L )
 },
 \label{AL-seriesalt}
 \end{equation}

For large contact lengths $\tau_L \gg
1$, the MT correction reduces to the standard equation for the
one-dimensional wire
 \begin{equation}
 \delta I_{MT}
 =
 \frac{\pi}{4}
 e^2V
 \frac{1}{\tau}
 \left(
   \frac{1}{\gamma^{1/2}}
   -
   \frac{1}{\tau_L^{1/2}}
 \right).
 \label{MT-1D}
 \end{equation}
As well as the AL correction, the MT correction remains finite
at $T = T_c$ and diverges at $T = T_c^*$ according to the same
law 
 \begin{equation}
 \delta I_{MT}
 =
 \frac{1}{2\pi}
 \frac{e^2TV}{T - T_c^*}.
 \label{MT-Tc}
 \end{equation}
In Fig. 1 we plotted the temperature dependences of the MT and AL 
terms in dimentionless fluctuation conductances $S_{MT}$ and $S_{AL}$
 for different values of $\gamma$ (the ratio of the depairing rate $\Gamma$ 
and the Thouless energy $\epsilon_L$); here $S_{MT} = \delta I_{MT} / I_o  \, , \,  
S_{AL} = \delta I_{AL} / I_o \, , \, I_o = e^2 V T / \epsilon_L$.
As  may be seen from 
Fig. 1, the MT contribution dominates over the whole temperature range. 
The depairing rate $\Gamma$ can be determined from measurements of the 
fluctuation conductance.
\begin{figure}[h]
\centerline{\psfig{figure=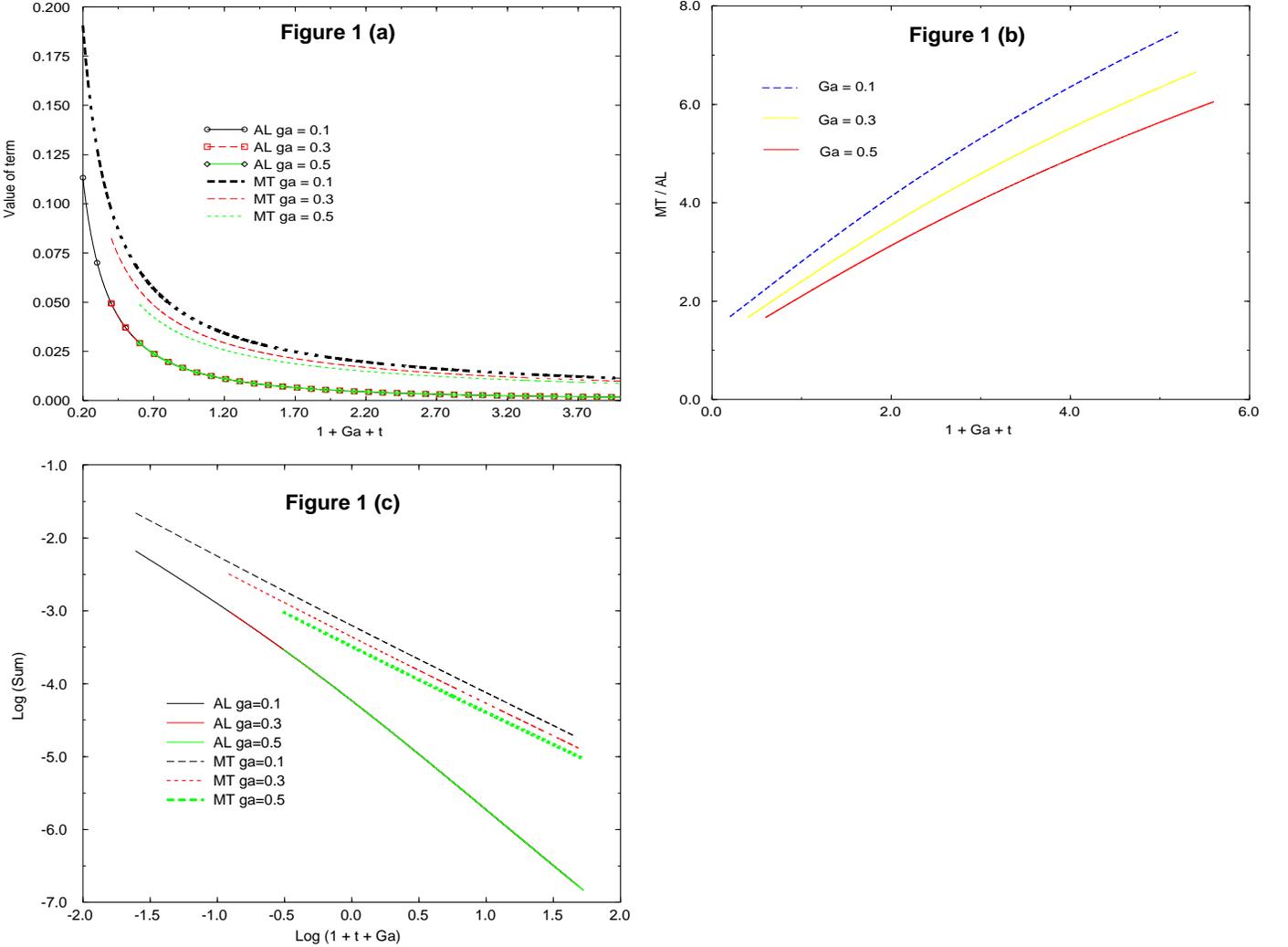,width=22cm,height=15cm}}
\caption{Figure 1 (a), shows the MT and the AL terms for several values of $\gamma$ (ga).
Figure 1 (b), shows the ration of MT/AL terms for several values of $\gamma$ (ga).
Figure 1 (c) is a Log-Log plot of Figure 1 (a).}
\label{}
\end{figure}
\section{ SNS CONTACT }

Consider a structure of similar geometry as the structures considered 
perviously but with a normal-metal
channel and superconducting electrodes.  The superconductor and
normal metal are characterized by the diffusion coefficients
$D_s$ and $D_n$ and by the phase-breaking rates $\Gamma_s$ and
$\Gamma_n$, respectively. In the case of SNS contacts, the
fluctuational Cooper pairs generated in the superconducting
electrodes can penetrate into the normal metal only a distance
shorter than $\xi_0$ as the BCS coupling constant $\lambda$
is zero in the normal metal.  However, as well as in the
superconducting state at $T < T_c$ \cite{Volkov-96},
\cite{Nazarov}, they can affect the
conduction of normal electrons at distances much larger than the 
phase-breaking length $L_{\varphi}$. In this case,
the conductance of the contact depends on the geometry of the 
electrodes even though their dimensions are much larger than the
transverse dimensions of the contact.

In view of this reasoning, the AL correction is negligible in
SNS contacts. The MT correction is obtained by averaging Eqn.
(\ref{MT-basic}) over the contact length, then integrating with
respect to ${\bf r}_1$, ${\bf r}_2$, and ${\bf r'}$ 
over the bulk of both electrodes, each giving an
independent contribution to the conductance of the contact.
Hence in Eqn. (\ref{MT-basic}), one may use the expressions for
$K^{R(A)}$ in a bulk homogeneous superconductor:
 \begin{equation}
 K^R(\omega, {\bf r}, {\bf r'})
 =
 -\frac{i}{\Omega}
 \sum_{\bf q}
 \frac
 {
   \exp[i{\bf q}({\bf r} - {\bf r'})]
 }
 {
   \omega
   +
   i[D_s q^2 + \tau T_c + \Gamma_s]
 },
 \label{KR-bulk}
 \end{equation}
where $\Omega$ is the normalization volume.

First consider the contribution from the left electrode. Assume
that the origin coincides with the left end of the contact. As
in the case of NSN contact, all the quantities inside
the channel depend only on the longitudinal coordinate $x$. As
the integral (\ref{MT-basic}) is dominated by $|{\bf r}_1|$ and
$|{\bf r}_2|$ of the order of $\xi(T)$, i.e., much larger than
the transverse dimensions of the contact, the quantity
$P^R(\epsilon, x, {\bf r})$, where $x$ is the coordinate of a
point inside the contact and $\bf r$ is the coordinate of a
point inside the left electrode, may be represented in the form
 \begin{equation}
 P^{R(A)}(\epsilon, x, {\bf r})
 =
 P_0^{R(A)}(\epsilon, -{\bf r})
 \,
 \psi(\pm\epsilon, x),
 \label{PR-product}
 \end{equation}
where $P_0^R(\epsilon, -{\bf r})$ is given by Eqn. (\ref{PR0})
and function $\psi$ is the solution of the equation
 $$
 D_n \frac{d^2\psi}{dx^2}
 +
 (i\epsilon - \Gamma_n)
 \psi(\epsilon, x)
 =
 0
 $$
with the boundary conditions $\psi(0) = 1$ and $\psi(L) = 0$.
Explicitly, it is given by the expression
 \begin{equation}
 \psi(\epsilon, x)
 =
 \frac
 {
 \sinh[\kappa(1 - x/L)]
 }
 {
 \sinh\kappa
 },
 \qquad
 \kappa(\epsilon)
 =
 (\gamma_n - i\epsilon/\epsilon_L)^{1/2}.
 \label{psi-explicit}
 \end{equation}
With these expressions and taking into account contributions
from both electrodes, Eqn. (\ref{MT-basic}) takes the form 
 \begin{equation}
 \delta I_{MT}
 =
 64e^2\epsilon_L TLSV
 \frac{1}{\Omega}
 \sum_{\bf q}
 \int\frac{d\omega}{2\pi}
 \,
 \frac{1}
 {
   \omega^2
   +
   (
     D_sq^2
     +
     \tau T_c
     +
     \Gamma_s
   )^2
 }
 \int\frac{d\epsilon}{2\pi}
 \,
 \frac{1}
 {
   4\epsilon^2
   +
   (
      D_s q^2
      +
      \Gamma_s
   )^2
 }
 \Phi(2\epsilon),
 \label{MT-SNS-1}
 \end{equation}
where
 $$
 \Phi(\epsilon)
 =
 \frac{1}{L}
 \int\limits_0^L dx
 \,
 \psi(\epsilon, x)
 \psi(-\epsilon, x)
 =
 \frac
 {
   (2\kappa_1)^{-1}
   \sinh(2\kappa_1)
   -
   (2\kappa_2)^{-1}
   \sin(2\kappa_2)
 }
 {
   \cosh(2\kappa_1)
   -
   \cos(2\kappa_2)
 },
 $$
 \begin{equation}
 \kappa_1 = {\rm Re} [\kappa(\epsilon)],
 \quad
 \kappa_2 = {\rm Im} [\kappa(\epsilon)].
 \label{Phi}
 \end{equation}
Integrating with respect to frequency $\omega$ in Eqn.
(\ref{MT-SNS-1}) gives
 \begin{equation}
 \delta I_{MT}
 =
 32e^2\epsilon_L TLSV
 \frac{1}{\Omega}
 \sum_{\bf q}
 \frac{1}
 {
     D_sq^2
     +
     \tau T_c
     +
     \Gamma_s
 }
 \int\frac{d\epsilon}{2\pi}
 \,
 \frac{1}
 {
   4\epsilon^2
   +
   (
      D_s q^2
      +
      \Gamma_s
   )^2
 }
 \Phi(2\epsilon).
 \label{MT-SNS-2}
 \end{equation}

To be specific, consider the case where the electrodes represent a
film of thickness $d_0 < \xi(T)$, then in this case the sum over
$\bf q$ may be replaced by the integral
 $$
 \frac{1}{\Omega} \sum_{\bf q}
 \to
 \frac{1}{d_0}
 \int\frac{d^2 q}{(2\pi)^2}.
 $$
Introducing the dimensionless integration variable $\theta =
2\epsilon/\epsilon_L$, one arrives at the following expression:
 $$
 \delta I_{MT}
 =
 \frac{4e^2}{\pi^2}
 \frac{TLS}{D_sd_0}
 V
 \int\limits_0^{\infty}
 \frac{d\theta}{\theta^2 + \tau_L^2}
 \left[
    \ln
    \left(
       \frac
       {
         \sqrt{\theta^2 + \gamma_s^2}
       }
       {
         \tau_L + \gamma_s
       }
    \right)
    +
    \frac{\tau_L}{\theta}
    \arctan
    \left(
       \frac{\theta}{\gamma_s}
    \right)
 \right]
 \frac
 { 
   \theta_1^{-1}\sinh\theta_1
   -
   \theta_2^{-1}\sin\theta_2
 }
 {
   \cosh\theta_1
   -
   \cos\theta_2
 },
 $$   
 \begin{equation}
 \theta_1
 =
 2^{1/2}
 \left(
    \gamma_n
    +
    \sqrt{\gamma_n^2 + \theta^2}
 \right)^{1/2},
 \quad
 \theta_2
 =
 2^{1/2}
 \left(
    \sqrt{\gamma_n^2 + \theta^2}
    -
    \gamma_n
 \right)^{1/2}.
 \label{MT-2D}
 \end{equation}

 Assume for simplicity that the depairing rates in the
normal and superconducting metals are equal ($\gamma_s =
\gamma_n = \gamma$). First consider the case of a very short
contact, $\gamma \ll \tau_L \ll 1$. In this case, the integral
(\ref{MT-2D}) is dominated by $\theta \sim \tau_L$, so the
last factor in the integrand may be set equal to 1/3. This
yields 
 \begin{equation}
 \delta I_{MT}
 =
 \frac{1}{12}
 \frac{D_n}{D_s}
 \frac{e^2}{d_0\tau}
 \ln
 \left(
   1 + \frac{\tau T_c}{\Gamma}
 \right)
 \frac{SV}{L}.
 \label{SNS-short}
 \end{equation}

To within the factor $D_n/D_s$ and a numerical coefficient, the
correction to the conductivity of the contact material is equal
to the MT conductivity of the electrodes. A similar result was
previously obtained by Zaitsev \cite{Zaitsev-84} for short ScS
contacts. 

Consider now the case where the contact is of intermediate length, $\gamma
\ll 1 \ll \tau_L$, one obtains with logarithmic accuracy
 \begin{equation}
  \delta I_{MT}
 =
 \frac{1}{12}
 \frac{D_n}{D_s}
 \frac{e^2}{d_0\tau}
 \ln(1/\gamma)
 \frac{SV}{L}.
 \label{SNS-medium}
 \end{equation}

This expression differs from that for the short contact in that
the quantity $(8/\pi)(T - T_c)$ in the logarithm is
replaced by the Thouless energy $\epsilon_L$.

Lastly, consider the case of a long contact, $1 \ll \gamma \ll
\tau_L$. In that case, the integral (\ref{MT-2D}) is dominated
by $\theta \gg 1$, so the second factor in the integrand may be
approximated by $(2\theta)^{-1/2}$, yielding
 \begin{equation}
 \delta I_{MT}
 =
 \frac{1}{2}
 \frac{D_n}{D_s}
 \frac{e^2}{d_0\tau}
 \gamma^{-1/2}
 \frac{SV}{L}.
 \label{SNS-long}
 \end{equation}

The physical meaning of this result is that the superconducting
correlations that result in the MT correction penetrate into the
contact over the length $L_{\varphi} \sim \gamma^{-1/2}L \ll L$.

Note that unlike the case of NSN contact, the correction to the
total current is proportional to the cross-sectional area of the
contact.

\bigskip
\section{ Conclusions }
The fluctuation conductivities of NSN and SNS contacts of arbitrary 
lengths have been calculated. We have established that the fluctuation 
conductivity in NSN contacts consists of contributions from both the 
Maki - Thompson and Aslamasov - Larkin terms. The MT contribution 
dominates over the whole temperature range. However, near the 
renormalized 
critical temperature $T_c^*$  the ratio of the MT and AL terms does not 
contain any parameters, and equals about 1.56 .

In SNS contacts the AL contribution is absent. The increase in
the conductivity 
due to fluctuations is caused by the anomalous MT term containing the 
product of retarded and advanced functions (see Eqn. (\ref{MT-basic})).

The superconducting fluctuations modiify the density of states
and decrease the
conductivity. In the case of NSN and SNS contacts analyzed by us, this 
decrease is small (i.e. it does not diverge as $T$ tends to  $T_c^*$). 
However the correction to the conductivity due to the decrease of the 
density of states is essential in SNS contacts at  $T < T_c^*$, this 
leads to the reentrant behaviour of the conductance 
\cite{Volkov-96},\cite{Nazarov},
\cite{Moriond}, and is also essential in tunnel SIS junctions \cite{Kulik} 
and in layered superconductors \cite{Volkov-93},\cite{ioffe}

\bigskip
\section{ Acknowledments }

This work was supported by the Russian Foundation for Basic
Research, grant 96-02-16663-a,the Russian Superconductivity
Program, grant no. 96053, the Royal Society, and by the CRDF, grant no. RP1 - 165. We would also 
like to thank C. J. Lambert for his attention and useful
 suggestions to this work.


\begin{references}

\bibitem{Petrashov} V.T. Petrashov, V.N. Antonov, P. Delsing,
and T. Claeson, Phys. Rev. Lett. {\bf 70}, 347 (1993); Phys.
Rev. Lett. {\bf74}, 5268 (1995).

\bibitem{Pothier} H. Pothier, S. Gueron, D. Esteve, and M.H.
Devoret, Phys. Rev. Lett. {\bf 73}, 2488 (1994).

\bibitem{Vedgar} P.G.N. de Vedgar, T.A. Fulton, W.H. Mallison,
and R.E. Miller, Phys. Rev. Lett. {\bf 73}, 1416 (1994).

\bibitem{Dimoulas} H. Dimoulas, J.P. Heida, B.J. van Wees, T.M.
Klapwijk, W. van der Graaf, and G. Borghs, Phys. Rev. Lett. {\bf
74}, 602 (1992).

\bibitem{Courtois} H. Courtois, Ph. Grandit, D. Maily, and B.
Pannetier, Phys. Rev. Lett. {\bf 76}, 130 (1996).

\bibitem{Volkov-96} A.F. Volkov, N. Allsopp, and C.J. Lambert,
J. Phys.: Condens. Matter {\bf 8}, 45 (1996).

\bibitem{Nazarov} Yu. V. Nazarov and T.H. Stoof, Phys. Rev.
Lett. {\bf 76}, 823 (1996).

\bibitem{Moriond} A.F. Volkov and V.P. Pavlovskii, in {\it
Correlated Fermions and Transport in Mesoscopic Systems}, Proc.
XXXI Moriond Conf., Les Arcs, 1996, p. 267.

\bibitem{Kulik} I.O. Kulik, Sov. Phys. JETP {\bf 32}, 510 (1971).

\bibitem{Zaitsev-84} A. V. Zaitsev, Sov. Phys. Solid State {\bf
26}, 1619 (1984).

\bibitem{Larkin-73} A.I. Larkin and Yu.N. Ovchinnikov, J. Low
Temp. Phys. {\bf 10}, 407 (1973).

\bibitem{Volkov-93} A.F. Volkov, Phys. Lett. {\bf A 175}, 445
(1993); Solid State Commun. {bf 88}, 715 (1993).

\bibitem{Larkin-86} A.I. Larkin and Yu.N. Ovchinnikov, in {\it
Nonequilibrium Superconductivity}, Eds. D.N. Langenberg and A.I.
Larkin, Elsevier, Amsterdam, 1986, p. 493.

\bibitem{Aslamazov} L.G. Aslamazov and A.I. Larkin, Sov. Phys.
Solid State {\bf 10}, 875 (1968).

\bibitem{Maki} K. Maki, Progr. Theor. Phys {\bf 39}, 897 (1968).

\bibitem{Thompson} R.S. Thompson, Phys. Rev. {\bf B1}, 327
(1970). 

\bibitem{ioffe} A.I. Larkin, A.Varlamov, and L.Yu, Phys. Rev. {\bf B47}, 8936
(1993). 

\end{references}
\end{document}